\documentclass{iopart}

\usepackage{graphics}
\usepackage{graphicx}
\usepackage{subfigure}
\usepackage{color}
\usepackage{epsfig}
\begin{document}

\title{A study of uranium-based multilayers: I. Fabrication and structural characterisation}

\author{R Springell$^{1,2}$, S W Zochowski$^{1}$, R C C Ward$^{3}$, M R Wells$^{3}$, S D Brown$^{2,4}$, L Bouchenoire$^{2,4}$, F Wilhelm$^{2}$, S Langridge$^{5}$, W G Stirling$^{2,4}$ and G H Lander$^{6}$}

\address{$^{1}$Department of Physics and Astronomy, University College London, London WC1E 6BT, UK}
\address{$^{2}$European Synchrotron Radiation Facility, BP220, F-38043 Grenoble Cedex 09, France}
\address{$^{3}$Clarendon Laboratory, University of Oxford, Oxford OX1 3PU, UK}
\address{$^{4}$Department of Physics, University of Liverpool, Liverpool L69 7ZE, UK}
\address{$^{5}$ISIS, Rutherford Appleton Laboratory, Chilton, Oxfordshire OX11 0QX, UK}
\address{$^{6}$European Commission, JRC, Institute for Transuranium Elements, Postfach 2340, Karslruhe, D-76125, Germany}
\ead{ross.springell@esrf.fr}
\begin{abstract}

This paper addresses the structural characterisation of a series of
U/Fe, U/Co and U/Gd multilayers. X-ray reflectivity has been
employed to investigate the layer thickness and roughness parameters
along the growth direction and high-angle diffraction measurements
have been used to determine the crystal structure and orientation of
the layers. For the case of uranium/transition metal systems, the
interfaces are diffuse ($\mathrm{\sim17\AA}$) and the transition
metals are present in a polycrystalline form of their common bulk
phases with a preferred orientation along the closest packed planes;
Fe, bcc (110) and Co, hcp (001), respectively. The uranium is
present in a poorly crystalline orthorhombic, $\mathrm{\alpha}$-U
state. In contrast, the U/Gd multilayers have sharp interfaces with
negligible intermixing of atomic species, and have a roughness,
which is strongly dependent on the gadolinium layer thickness.
Diffraction spectra indicate a high degree of crystallinity in both
U and Gd layers with intensities consistent with the growth of a
novel hcp U phase, stabilised by the hcp gadolinium layers.

\end{abstract}

\maketitle

\section{Introduction}

The properties of a material can differ greatly from the bulk when
reduced in size into the nanometre regime. The fabrication of
multilayers results in the juxtaposition of different elements in
systems where the interfacial regions comprise a substantial part of
the whole sample, producing interesting electronic and magnetic
effects. Varying the structural composition of the different
elements can be used to manipulate the electronic and magnetic
behaviour of the respective constituents \cite{Bland}. The use of
the actinide element uranium in such systems can be used to
investigate effects arising from the unpaired $\textit{5f}$
electrons, which exhibit strong hybridisation with other electronic
states in uranium compounds \cite{Severin}. This tendency for
$\textit{5f}$ hybridisation could lead to exotic properties in
multilayers containing uranium.

The polarisation of uranium was reported in a study of the UAs/Co
multilayer system \cite{Fumagalli,Kernavanois}, where the proximity
of the amorphous ferromagnetic UAs compound to the transition metal
(TM) ferromagnet, Co, resulted in a large magneto-optical Kerr
effect from the uranium \cite{Fumagalli2}. The first reports of
multilayers including uranium in its elemental form discuss the
proximity effects of the transition metals Co \cite{Rosa} and Fe
\cite{Beesley1}. Our group has carried out a series of experiments
on U/Fe multilayers \cite{Beesley1,Beesley2}. These papers discuss
the fabrication and characterisation of a series of samples, using a
combination of X-ray reflectivity (XRR), X-ray diffraction (XRD),
M\"{o}ssbauer spectroscopy, bulk magnetisation and polarised neutron
reflectivity (PNR) techniques. Since these studies, modifications to
the sputtering apparatus at the Clarendon laboratory, Oxford, have
improved the control of the sputtering rates and the inclusion of a
third sputter-gun has allowed the growth of buffer and capping
layers to seed crystalline assembly of the bilayers and prevent
oxidation of the multilayer stack. Recent measurements of the X-ray
magnetic circular dichroism (XMCD) at the U M-edges have probed the
electronic behaviour of the U $\textit{5f}$ states in U/Fe
multilayers \cite{Wilhelm} and confirmed earlier X-ray resonant
magnetic reflectivity (XRMR) measurements \cite{Brown}, which
demonstrate that a polarisation of the uranium $\textit{5f}$
electrons occurs in the U/Fe multilayers as a result of the
hybridisation of (U)$\textit{5f}$-$\textit{3d}$(Fe) electrons.

In the present articles (I and II), we report on a new series of
U/Fe multilayers and extend our study to the transition-metal U/Co
system. In addition, the fabrication and characterisation of a
series of U/Gd multilayers is also described. In order to understand
the growth mechanisms and structural properties of this range of
systems it is helpful to recall the sizes of the respective atoms.
The atomic volumes of Fe and Co are $\mathrm{\sim12\AA^{3}}$,
whereas that of U is $\mathrm{\sim21\AA^{3}}$, assuming the latter
to be in the room temperature, ambient form of the alpha phase,
which has an orthorhombic crystal structure. The resulting mismatch
in one dimension is $\mathrm{\sim20\%}$ and would result in a
considerable compressive strain on the uranium. In contrast, the
atomic volume of Gd is $\mathrm{\sim33\AA^{3}}$, giving a length
mismatch of $\mathrm{\sim14\%}$ and a strain that is clearly in the
opposite sense to that found when using $\textit{3d}$ transition
metal elements. These strains could result in significantly
different structural and magnetic properties between the U/TM and
U/Gd multilayers.

\section{Fabrication}

Multilayers were fabricated using a three-gun, dc magnetron
sputtering assembly in a loadlocked growth chamber operating at UHV
base pressure (5$\mathrm{\times10^{-10}}$mbar). Substrates were
single-crystal sapphire plates, which were epi-polished parallel to
the ($\mathrm{11\bar{2}0}$) plane. A 50$\mathrm{\AA}$ thick niobium
buffer layer was used to seed crystalline growth of the bilayers. Nb
has a body centred cubic (bcc) crystal structure and is expected to
exhibit [110] preferred orientation on sapphire
($\mathrm{11\bar{2}0}$) when deposited at ambient temperature (it is
fully epitaxial at elevated temperature). A similar Nb layer was
used as a capping layer to prevent atmospheric attack of the
multilayer after growth. A study of epitaxial (110) Nb films on
sapphire has found that a stable layer of
$\mathrm{Nb_{2}O_{5}\sim20\AA}$ thick is formed, which provides
effective long-term passivation \cite{Hellwig}.

Sputtering was carried out in a (flowing) argon pressure of
$\mathrm{5\times10^{-3}}$ mbar, and a growth rate $\mathrm{\sim1\AA
s^{-1}}$ was employed for each element. Precise deposition rates
were determined from measurement of calibration samples, by
comparison of experimental and calculated X-ray reflectivity
profiles. The majority of samples were grown at ambient temperature,
although a substrate heater was available to investigate the effects
of elevated temperature on selected samples.

\section{Structural Characterisation}

Series of U/Fe, U/Co and U/Gd samples were made in order to study
the structural properties systematically as a function of the layer
thicknesses of the respective elements, and to contrast and compare
trends between the systems. In this paper the composition of the
multilayers is represented as $\mathrm{[A_{X}/B_{Y}]_{Z}}$, where A
and B represent the elements comprising the bilayer system of
interest, X and Y are the respective layer thicknesses
$\mathrm{(\AA)}$ as determined by X-ray reflectivity and Z is the
number of bilayer repeats. In this case, element A is uranium and
element B represents the ferromagnet, iron, cobalt or gadolinium.
The samples were grown with layer thicknesses in the ranges
$5<\mathrm{t_{U}(\AA)}<90$ and $10<\mathrm{t_{B}(\AA)}<80$. The
X-ray reflectivity technique was employed to investigate the
composition of the multilayers, in terms of the layer thickness and
interface roughness values. High-angle X-ray diffraction
measurements were used to investigate the crystal structures of the
respective elements within the layers.

\subsection{X-ray reflectivity}

\subsubsection{Experimental method}

X-ray reflectivity scans were carried out on a Philips
diffractometer at the Clarendon Laboratory, Oxford. This
diffractometer was optimised for low-angle diffraction measurements.
A copper anode tube source provided Cu K$\alpha$ X-rays of
wavelength 1.54$\mathrm{{\AA}}$, selected by a germanium
monochromator and attenuated by a nickel foil to avoid detector
saturation and to reduce Cu K$\beta$ background. The scans were
taken in a specular geometry with the scattering vector normal to
the sample surface.

The data were fitted to simulations of the reflected intensity,
based on a matrix method of interferometry that reduces to Parratt's
recursive method \cite{Parratt}. The interface regions were modeled
by a method proposed by $\mathrm{N\acute{e}vot}$ and Croce
\cite{Nevot}, which was later adapted by others \cite{Sinha, Stoev,
Nevot2}, which treats the layer roughnesses as a variation of the
index of refraction. The simulations and fitting routines are part
of the xPOLLY programme \cite{XPolly}. The reflectivity is
calculated by a set of input parameters, including the anomalous
scattering factors of the respective materials at the energy of the
incident photons, the density $(\mathrm{atoms}/\mathrm{{\AA}^{3}})$,
the layer thickness $(\mathrm{{\AA}})$ and the rms roughness
$(\mathrm{{\AA}})$. All of these values can be varied, although in
practice the scattering factors were kept constant and the
structural parameters varied. The initial structural models consist
of a substrate, Nb buffer layer, repeated bilayer and an oxidised
capping layer. Complexity can then be introduced by stratifying the
bilayer to account for regions of reduced density at the interfaces,
where strain, caused by lattice mismatches between relevant species,
can produce defects affecting the crystal structure of the layers.

This technique provides an excellent measure of the bilayer
thickness, but is limited in its sensitivity to the relative
thicknesses of individual layers. Good fits to the data could be
produced by simulations that varied in individual layer thickness by
several angstroms. For this reason, the reflectivity was not
considered in isolation, but consistency was maintained by
consideration of the growth parameters and results from XRD, PNR and
SQUID magnetometry measurements \cite{UGdpaperII}.

\subsubsection{Results}

The results are presented for the specular reflectivity of U/Fe,
U/Co and U/Gd systems respectively. The normalised reflected
intensity is plotted against the wave-vector momentum transfer, Q
($\mathrm{\AA}^{-1}$), normal to the sample surface, where $2\theta$
is the scattering angle, $Q=2k\sin\theta$ and the wavevector
$k=\frac{2\pi}{\lambda}$. This scattering geometry probes the
reflected intensity as a function of depth, where the X-rays are
sensitive to the electron density profile.

\begin{figure}[htbp]
\center
\mbox{\subfigure[$\mathrm{SN71-[U_{9}/Fe_{34}]_{30}}$]{\epsfig{figure=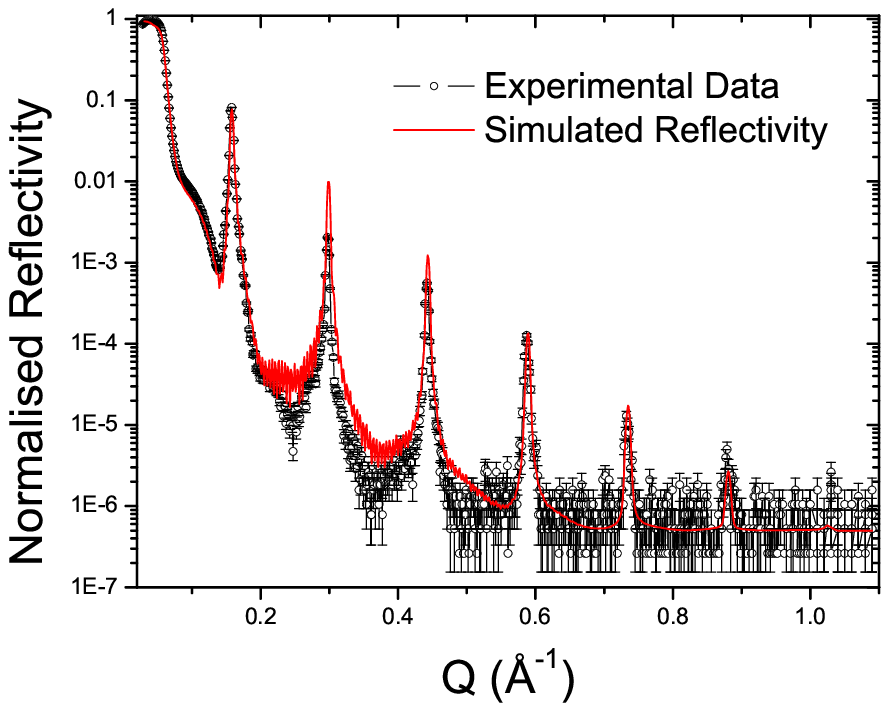,width=0.45\textwidth,bb=10
10 280 225,clip}}\quad
\subfigure[$\mathrm{SN74-[U_{32}/Fe_{27}]_{30}}$]{\epsfig{figure=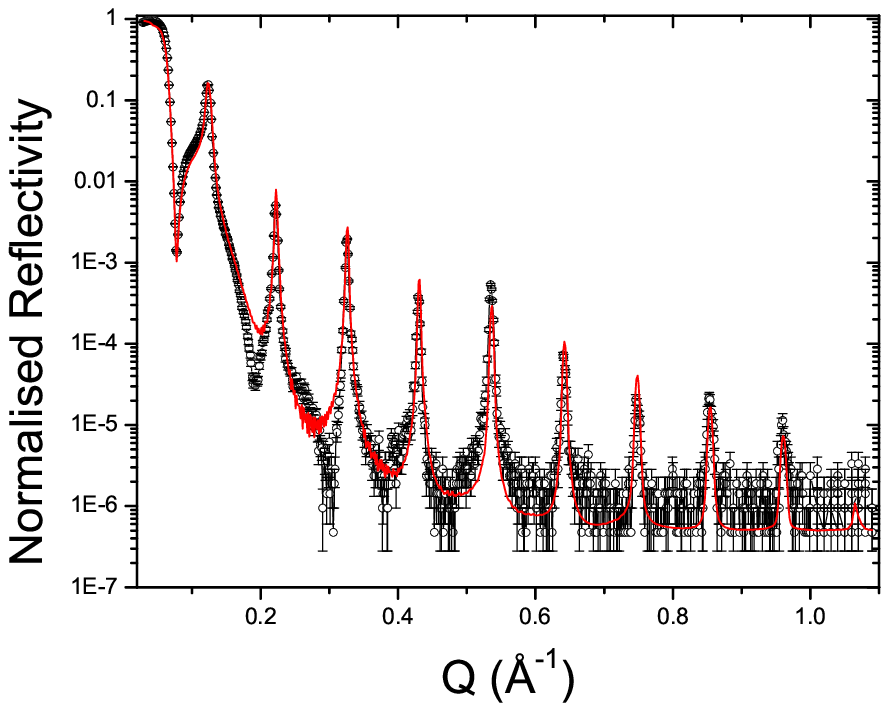,width=0.45\textwidth,bb=10
10 280 225,clip}}}
\mbox{\subfigure[$\mathrm{SN108-[U_{27.5}/Co_{27.5}]_{20}}$]{\epsfig{figure=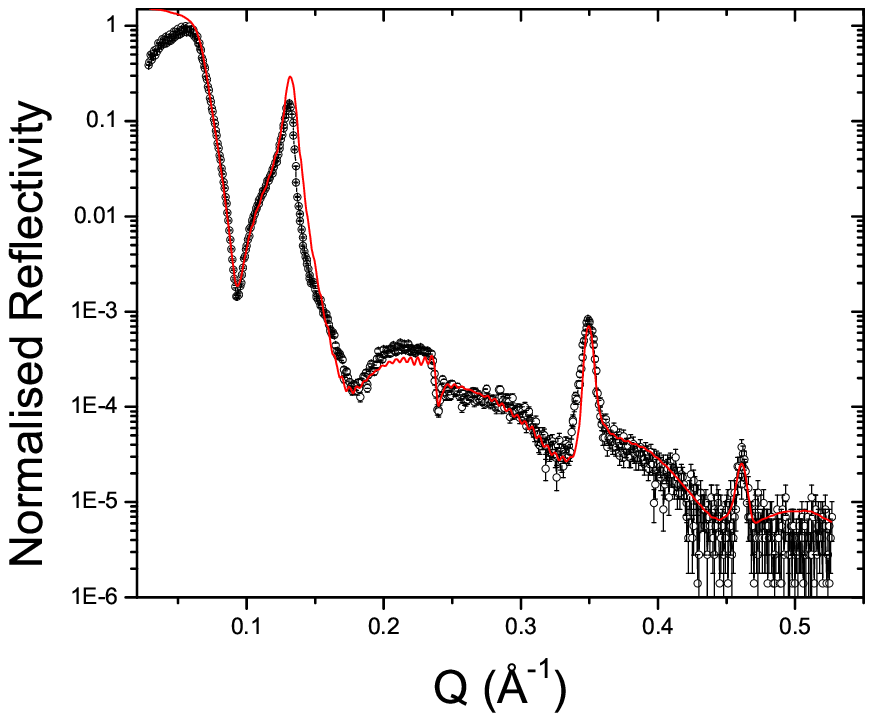,width=0.45\textwidth,bb=10
10 280 225,clip}}\quad
\subfigure[$\mathrm{SN112-[U_{19}/Co_{19.2}]_{20}}$]{\epsfig{figure=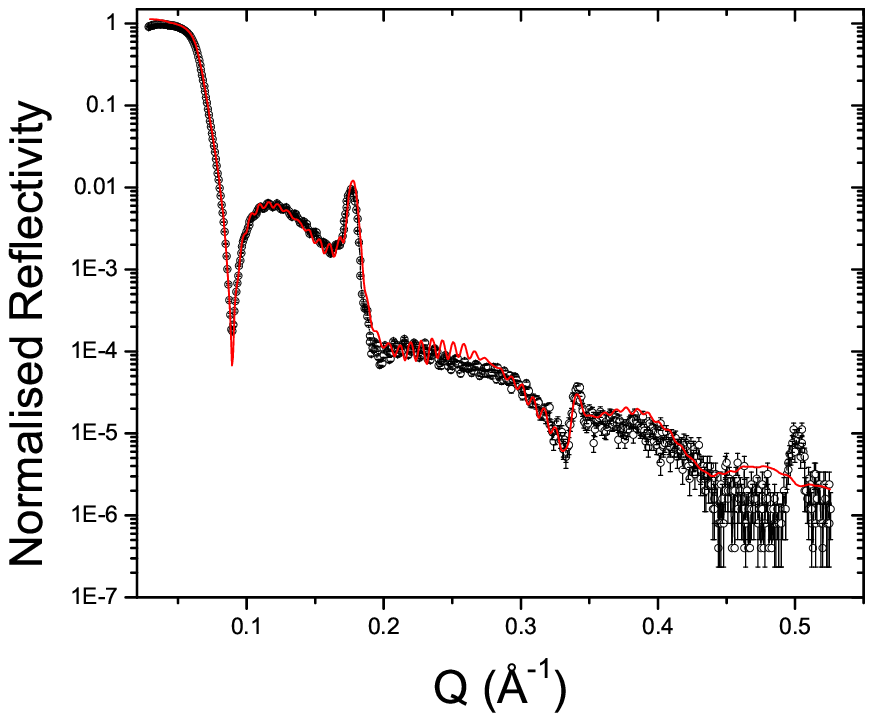,width=0.45\textwidth,bb=10
10 280 225,clip}}}\caption{\label{UFeCoxref}U/Fe and U/Co X-ray
reflectivity spectra taken at room temperature, using a Cu
K$\mathrm{\alpha}$ X-ray source. Calculated reflectivity curves
(shown in red) were fitted to the data using the xPOLLY programme.
Layer thicknesses are as fitted; we note that the precision is of
order $\mathrm{1\AA}$ (see text).}
\end{figure}

Figure \ref{UFeCoxref} shows example X-ray reflectivity spectra for
U/Fe ((a) and (b)) and U/Co samples ((c) and (d)). The work of
Beesley et al. \cite{Beesley1} \cite{Beesley2}, based on conclusions
made from M\"{o}ssbauer spectroscopy measurements, stratified the
iron layers into three components: a highly textured (110) bcc
crystalline component with a magnetic moment value close to that in
bulk Fe, a crystalline component with a reduced moment, and a region
of amorphous, non-magnetic iron ($\mathrm{\sim12\AA}$) located at
the interfaces. Further analysis of M\"{o}ssbauer data and a
consideration of the mechanism for the existence of non-magnetic Fe
has led to a modified view of the iron layer structure. The
non-magnetic iron component detected by M\"{o}ssbauer spectroscopy
indicates that the Fe atoms are in an environment that leads to
equally populated spin-up and spin-down bands. This situation is
most likely to occur in an alloy, formed by interdiffusion at the
U/Fe interfaces. We now believe that the Fe layer is best described
thus: $\mathrm{UFe_{alloy}|Fe_{amorphous}|Fe_{bcc}|FeU_{alloy}}$. In
reality, distinct boundaries will not exist between these respective
components.

\begin{figure}[htbp]
\center
\mbox{\subfigure[$\mathrm{SN65-[U_{26}/Gd_{76}]_{20}}$]{\epsfig{figure=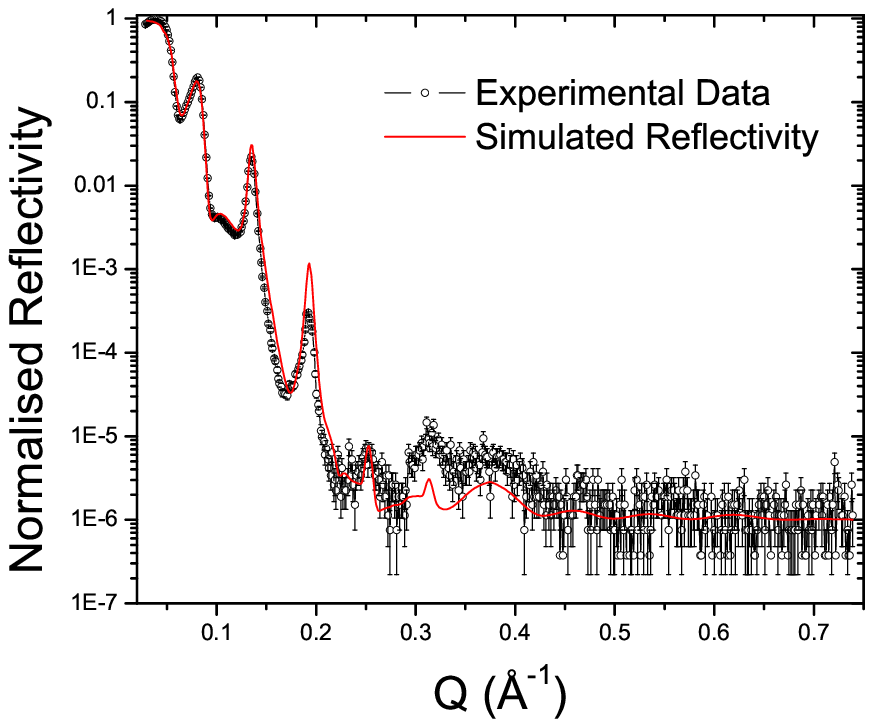,width=0.45\textwidth,bb=10
10 280 225,clip}}\quad
\subfigure[$\mathrm{SN68-[U_{89}/Gd_{20}]_{20}}$]{\epsfig{figure=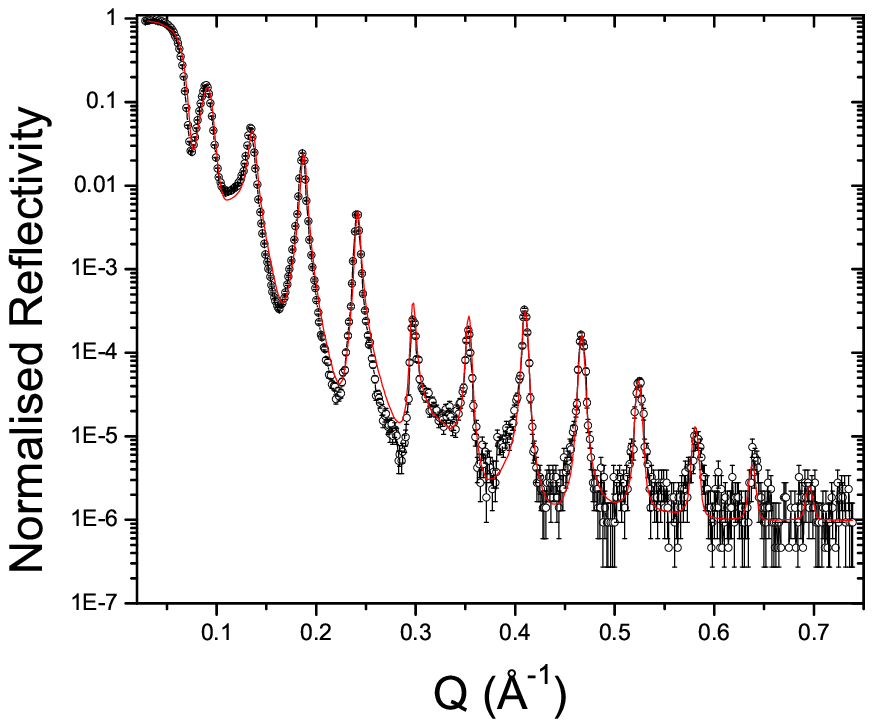,width=0.45\textwidth,bb=10
10 280 225,clip}}}
\mbox{\subfigure[$\mathrm{SN63-[U_{26}/Gd_{33}]_{20}}$]{\epsfig{figure=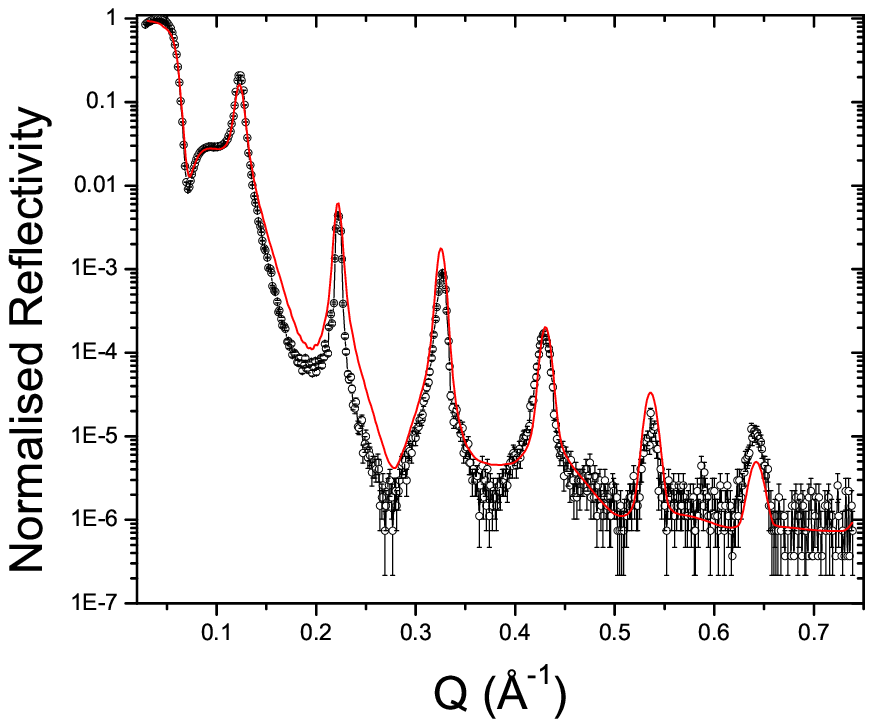,width=0.45\textwidth,bb=10
10 280 225,clip}}\quad
\subfigure[$\mathrm{SN123(4)-[U_{11.1}/Gd_{24}]_{20}}$]{\epsfig{figure=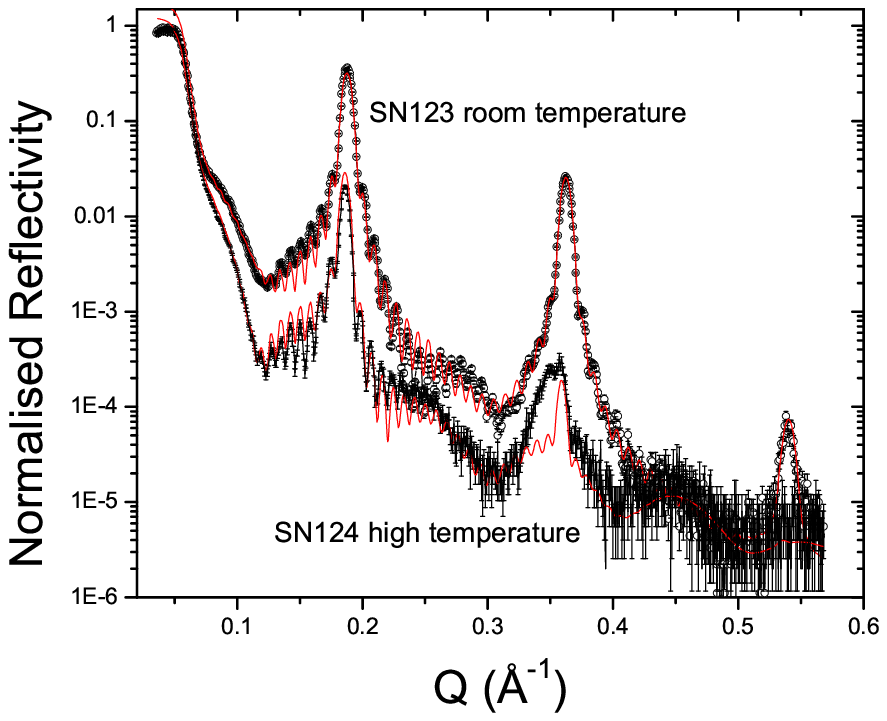,width=0.45\textwidth,bb=10
10 280 225,clip}}}
\mbox{\subfigure[$\mathrm{SN137-[U_{28.2}/Gd_{19.5}]_{30}}$]{\epsfig{figure=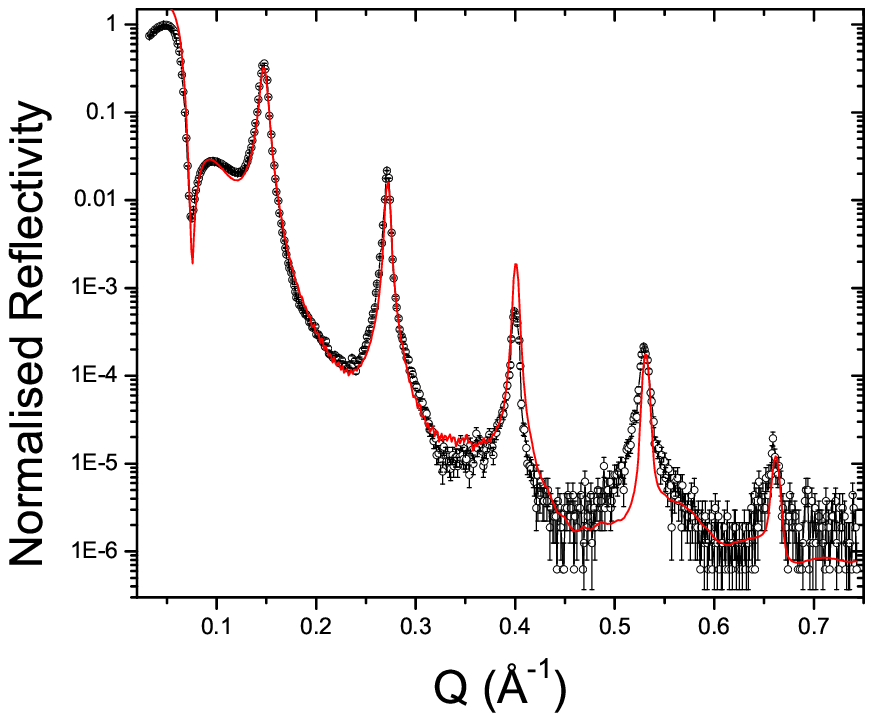,width=0.45\textwidth,bb=10
10 280 225,clip}}\quad
\subfigure[$\mathrm{SN139-U_{3.5\%}Gd}$]{\epsfig{figure=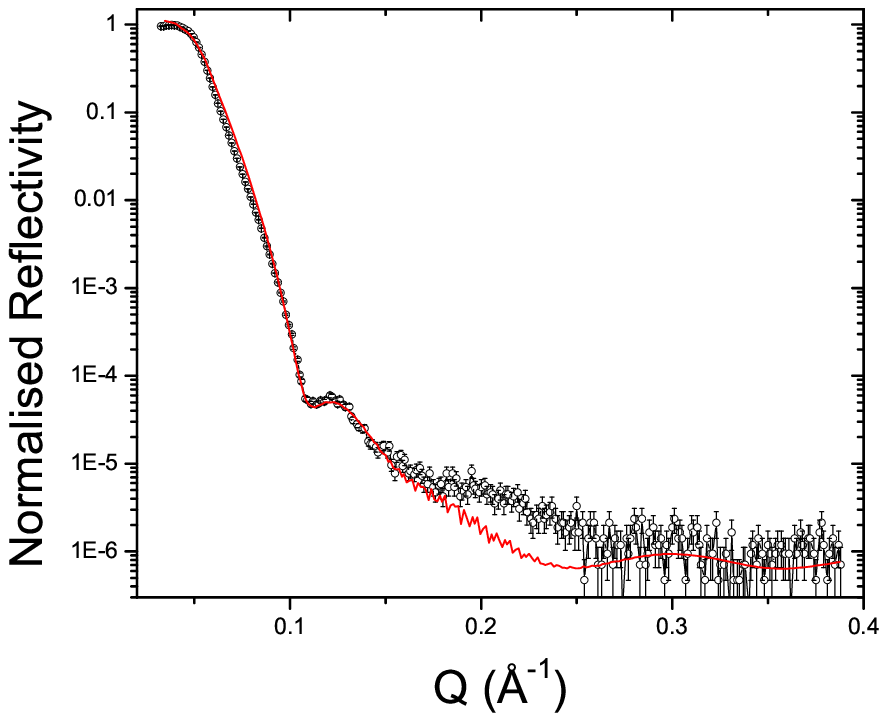,width=0.45\textwidth,bb=10
10 280 225,clip}}}\caption{\label{UGdxref}U/Gd X-ray reflectivity
spectra taken at room temperature, using a Cu K$\mathrm{\alpha}$
X-ray source. Calculated reflectivity curves (shown in red) were
fitted to the data using the xPOLLY programme \cite{XPolly}. Note
that the sample SN139 shown in panel (f) is not a multilayer, but a
sputtered alloy of $\mathrm{\sim3.5\%}$ U in Gd.}
\end{figure}

The total bilayer repeat distance can be determined with a precision
of $\mathrm{0.1\AA}$, but the individual layer thicknesses could not
be so well defined. However, restrictions were also fixed on these
values based on the sputtering times and known calibrations. The
roughness values alter the Bragg peak intensities and increases
greater than $\mathrm{\sim1\AA}$ in the rms roughness per layer can
result in major reductions in the reflected intensity. However, it
is not possible to distinguish between interdiffusion and roughness
at the interface since they produce equivalent effects in the
specular reflected intensity. More information on the nature of the
interfacial structure can be obtained by a combination of high-angle
X-ray diffraction and X-ray reflectivity measurements.

The U/Fe samples analyzed previous to this work using X-ray
reflectivity \cite{Beesley1} were grown on glass substrates and did
not include either a buffer or a capping layer to prevent oxidation.
The use of an Nb capping layer reduces the complicated oxidation
profile through the multilayer stack to a $\mathrm{Nb/Nb_{2}O_{5}}$
surface layer. This gives a more simpler for the calculation of the
reflected intensity and one which includes a similar surface
contribution for all of the samples. For samples of similar
thickness the single crystal, optically flat, sapphire substrates
have reduced the respective roughness of the individual layers.

The reflectivity results for the U/Co samples were fitted by
separating the cobalt layers into two components; one reduced
density ($\mathrm{>90\%}$ of the bulk value) component of
$\mathrm{\sim15\AA}$ thickness and the remainder of the bulk Co
density, $\mathrm{\rho_{Co}=9\times10^{28}m^{-3}}$. This structural
profile was determined from the polarised neutron reflectivity and
bulk magnetisation measurements to be discussed in paper II
\cite{UGdpaperII} of this series. It was not possible to identify a
U-Co alloy region at the interface as indicated by M\"{o}ssbauer
measurements on U/Fe samples, although it is likely to be present
due to the similar atomic sizes of Fe and Co resulting in similar
interfacial strains. The majority of the features observed in the
reflected intensity have been reproduced in the simulations,
including the extinction of even order Bragg peaks in the case of
samples SN108 and SN112 in figure \ref{UFeCoxref} (c) and (d), where
the thicknesses $\mathrm{t_{U}}$ and $\mathrm{t_{Co}}$ are almost
equal.

Figure \ref{UGdxref} shows observed reflectivity spectra for a range
of U/Gd multilayers. Panels (a) and (b) have similar bilayer repeat
thicknesses, but vary in composition between thick Gd layers and
thick U layers respectively. The difference in the spectra is
striking; for large values of $\mathrm{t_{Gd}}$ (a) the reflected
intensity decreases rapidly as a function of Q
($\mathrm{\AA^{-1}}$), compared with the observation of well-defined
Bragg peaks over a wide Q-range in the reflectivity spectra of
samples with thick U layers, e.g. (b). The graphs (a), (c) and (e)
show the reflectivity curves for samples of decreasing Gd layer
thickness for almost constant values of $\mathrm{t_{U}}$ and
indicate a reduction in the bilayer roughness for thin Gd layers.
Figure \ref{UGdxref} (d) compares the observed reflectivity for two
like samples grown at different temperatures and (f) contrasts the
reflectivity spectrum observed for a U-Gd alloy sample with those of
U/Gd multilayers.

\subsubsection{Discussion}

The general good quality of multilayer samples in all cases is
supported by the form of the measured X-ray reflectivity profiles.
The relative growth properties of U/Fe samples grown on sapphire
substrates with niobium buffer and capping layers, can be compared
to those grown previously on glass with no buffer or capping layers
\cite{Beesley1}, by comparing relative thickness and roughness
parameters. The roughness of layers in the latter, although
$\mathrm{\sim1-2\AA}$ larger for samples of similar layer thickness,
are of approximately the same magnitude, indicating that the bilayer
growth mechanisms are the same in both cases and that the majority
of the roughness stems from the relative lattice mismatch and
crystalline nature of the respective species. The slightly reduced
roughness in the new samples can be understood as an effect of the
smooth substrate surface and low roughness value of the niobium
buffer layer.

Both U/Fe and U/Co samples were modeled by separating the
ferromagnetic layers into two components; one with a reduced density
($\mathrm{10\%}$ less than the bulk value), $\mathrm{\sim15\AA}$
thick, and the remainder of the layer with the bulk density. This
model is supported by results obtained in PNR, Mossbauer and SQUID
magnetometry \cite{Beesley1} \cite{Beesley2} measurements, discussed
in paper II \cite{UGdpaperII}, and can be understood by considering
the growth of layers with a large mismatch in lattice spacings,
$\mathrm{\sim20\%}$. The large strains and diffusion of the smaller
transition metal atoms into the uranium layers produce an alloyed
region at the interfaces. Growth of the ferromagnetic layers onto
these alloys produces an initial amorphous, noncrystalline form, but
as the layer thickness is increased the layer tends towards a bulk
crystalline state.

All uranium/gadolinium samples were modeled with a simple bilayer
structure, since magnetisation measurements \cite{UGdpaperII} have
not revealed the presence of any substantial 'dead' layer, requiring
a stratified density gadolinium layer. X-ray diffraction
measurements (described in section 3.2) indicate a much lower
lattice mismatch between U and Gd layers than that observed in the
transition metal systems, which could lead to a more coherent layer
by layer growth and therefore not require such a complex description
of the gadolinium layer structure. As shown in figure \ref{UGdxref}
and table \ref{roughnesses}, for thick uranium layers a large number
of Bragg peaks were observed over a wide Q range, characterised by a
low rms roughness of $\mathrm{\sim4\AA}$ per layer. For an
equivalent bilayer thickness, but with thick gadolinium layers, the
roughness was larger, possibly caused by a more columnar crystal
growth, resulting in a step-like roughness profile. The large
difference between the X-ray reflectivity in these two cases was not
apparent for similar situations in the U/Fe and U/Co systems.

\begin{table}[htbp]\caption{\label{roughnesses}Roughness values per
layer ($\mathrm{\AA\pm10\%}$) as a function of (a) gadolinium thickness and
(b) uranium layer thickness.} \centering
\subtable[]{\begin{tabular}{clccc}\br Sample & Composition & $\sigma_{U}$ & $\sigma_{Gd}$ & $\sigma_{av}$ \\
\mr
SN63 & $\mathrm{[U_{26}/Gd_{33}]_{20}}$ & 3.2 & 7.0 & 5.1 \\
SN64 & $\mathrm{[U_{26}/Gd_{54}]_{20}}$ & 7.0 & 7.0 & 7.0 \\
SN65 & $\mathrm{[U_{26}/Gd_{76}]_{20}}$ & 9.0 & 10.0 & 9.5 \\
\br\end{tabular}} \space\space
\subtable[]{\begin{tabular}{clccc}\br Sample & Composition & $\sigma_{U}$ & $\sigma_{Gd}$ & $\sigma_{av}$ \\
\mr
SN66 & $\mathrm{[U_{39}/Gd_{20}]_{20}}$ & 3.0 & 5.0 & 4.0 \\
SN67 & $\mathrm{[U_{63.5}/Gd_{20}]_{20}}$ & 3.2 & 5.5 & 4.4 \\
SN68 & $\mathrm{[U_{89}/Gd_{20}]_{20}}$ & 2.9 & 5.0 & 4.0 \\
\br\end{tabular}}
\end{table}

Tables \ref{roughnesses} (a) and (b) show the rms roughness values
for a selection of U/Gd samples; a set with constant
$\mathrm{t_{U}}$ and increasing $\mathrm{t_{Gd}}$ and a series with
constant $\mathrm{t_{Gd}}$ and varying $\mathrm{t_{U}}$. Average
roughness values are given in $\mathrm{\AA}$ for the uranium
($\sigma_{U}$) and gadolinium ($\sigma_{Gd}$) layers, and
$\sigma_{av}$ represents an average roughness per bilayer. Similar
to the determination of individual layer thicknesses from the
simulation of the X-ray reflectivity spectra, the individual layer
roughnesses were also difficult to distinguish precisely, although
the intensities were very sensitive to $\sigma_{av}$. However, the
vast majority of samples studied indicated larger roughness values
for the gadolinium layers than for the uranium. Table
\ref{roughnesses} (a) shows a near linear relationship between
$\mathrm{t_{Gd}}$ and $\sigma_{av}$, where thicker Gd layers result
in large rms roughness values. In contrast, table \ref{roughnesses}
(b) shows a practically constant $\sigma_{av}$ for a range of U
layer thicknesses.

\subsection{X-ray Diffraction}

The previous section has dealt with the use of X-rays to probe the
physical composition of the multilayers on length scales
$\mathrm{\sim10\rightarrow1000\AA}$, perpendicular to the plane of
the sample. It is also important, however, to be able to determine
the crystal structure and orientation of the respective layers and
various properties of the crystallites that have formed. A study of
this type gives insight into the growth mechanisms and interfacial
structure of the multilayer samples. X-ray diffraction is the most
commonly used and readily available tool for the investigation of
these properties.

Due to the likely lattice mismatch between the respective elements
in the case of U/TM and U/Gd systems the samples considered here are
likely to be composed mainly of polycrystalline layers with a
preferred orientation and a range of crystallite sizes.

\subsubsection{Results}

\begin{figure}[htbp]
\center
\mbox{\subfigure{\epsfig{figure=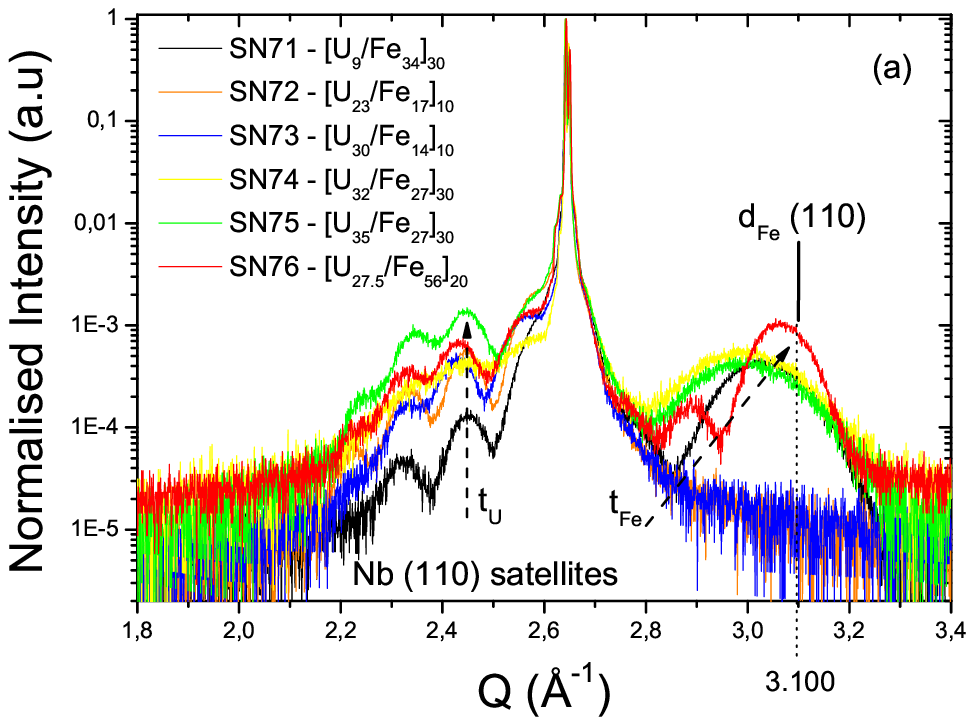,width=0.7\textwidth,bb=10
10 300 240,clip}}}
\mbox{\subfigure{\epsfig{figure=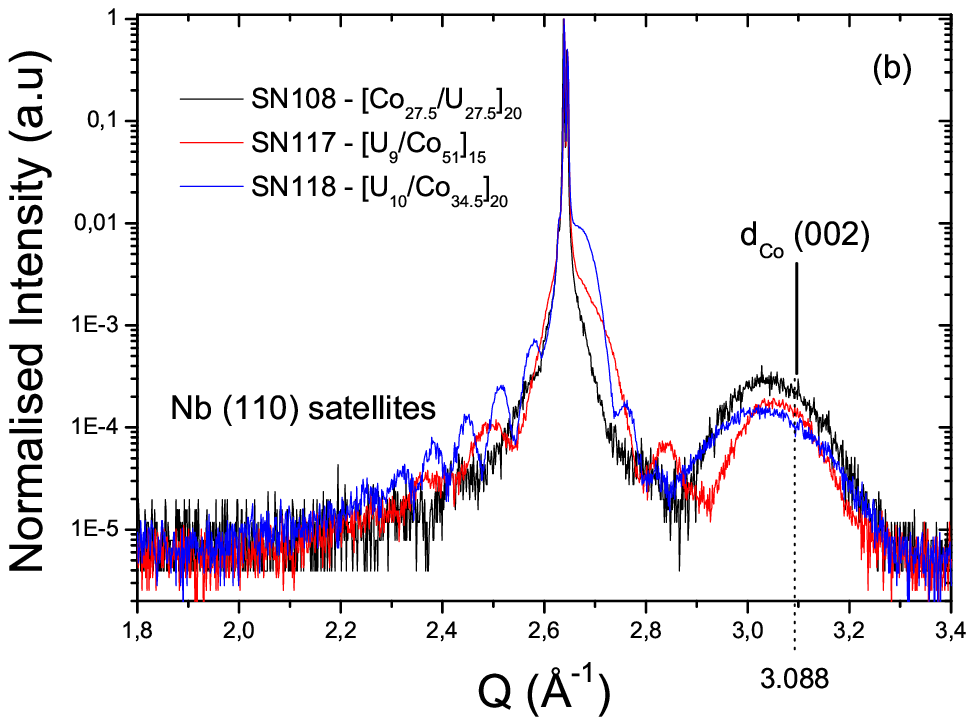,width=0.7\textwidth,bb=10
10 300 240,clip}}}\caption{\label{UFeCoxdif}Comparison of the X-ray
diffraction patterns close to the sapphire ($11\bar{2}0$) peak, for
selected U/Fe samples (upper panel) and selected U/Co samples (lower
panel). The arrows in panel (a)  indicate increasing U and Fe layer
thicknesses. The d-spacings of Fe (110) and Co(002) are indicated.}
\end{figure}

\begin{figure}[htbp]
\centering
\includegraphics[width=0.7\textwidth,bb=10 10 300 225,clip]{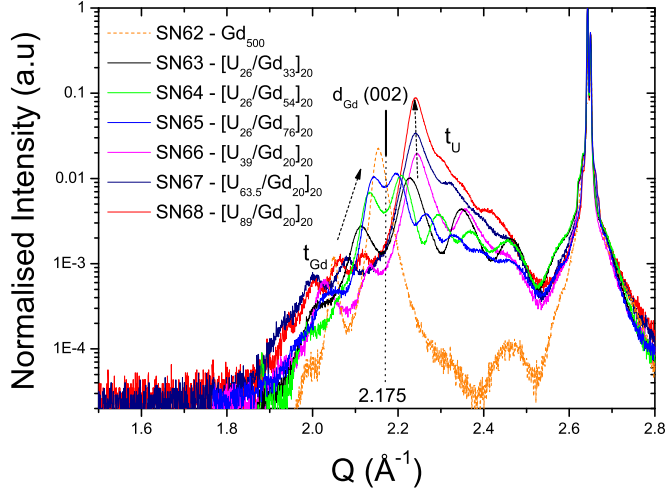}\caption
{\label{UGd1xdif}Comparison of the X-ray diffraction patterns close
to the sapphire ($11\bar{2}0$) peak, for a series of U/Gd samples.
The dashed orange peak represents the diffraction from a
$\mathrm{500\AA}$ thick sputtered Gd film, which results in a
lattice spacing, $\mathrm{d_{Gd}}$ of $\mathrm{2.92\AA}$. The
position of the bulk values is also noted.}
\end{figure}

\begin{figure}[htbp]
\centering
\includegraphics[width=0.7\textwidth,bb=10 10 300 225,clip]{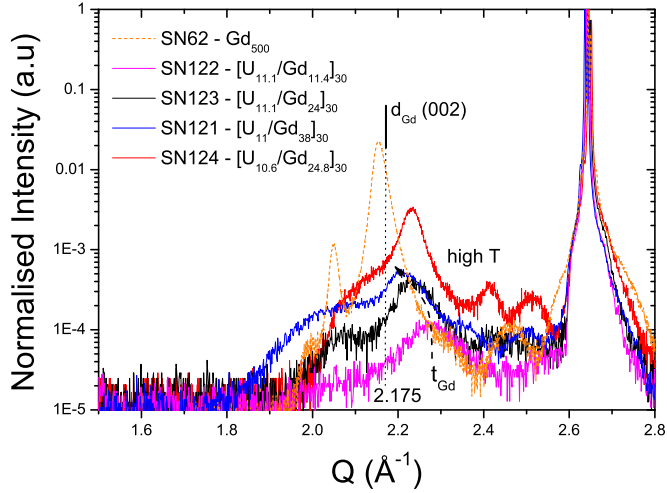}\caption
{\label{UGd2xdif}Comparison of the X-ray diffraction patterns close
to the sapphire ($11\bar{2}0$) peak, for a selection of U/Gd
samples. The dashed peak represents the diffraction from a
$\mathrm{500\AA}$ thick sputtered Gd film, which results in a
lattice spacing, $\mathrm{d_{Gd}}$ of $\mathrm{2.92\AA}$. The
position of the bulk gadolinium lattice spacing is also noted.}
\end{figure}

The results are presented for the X-ray diffraction in a
$\mathrm{\theta-2\theta}$ geometry for U/Fe, U/Co and U/Gd systems,
respectively. Summaries of the X-ray diffraction patterns for U/Fe
and U/Co systems are presented in figures \ref{UFeCoxdif} (a) and
(b), whereas figures \ref{UGd1xdif} and \ref{UGd2xdif} summarise the
series of U/Gd samples. The diffracted intensity is plotted against
the momentum transfer, Q ($\mathrm{\AA^{-1}}$), for each series of
samples in order to compare qualitatively structural variations of
the properties across the series. The intensity is normalised to
unity at the peak of the scattering from the sapphire substrate.

The upper panel of figure \ref{UFeCoxdif} shows a summary of the
X-ray diffraction patterns taken for the U/Fe series of samples. The
intense peak at $\mathrm{2.643\AA^{-1}}$ is due to the epitaxial
sapphire substrate and the satellite peaks that appear on the low
angle side of the substrate peak are a consequence of the
$\mathrm{\sim50\AA}$ thick niobium buffer layer. The
$\alpha$-uranium (110), (021) and (002) peaks were observed
previously in diffraction spectra of U/Fe multilayers grown on glass
\cite{Beesley1} and were positioned at $\mathrm{2.448\AA^{-1}}$,
$\mathrm{2.490\AA^{-1}}$ and $\mathrm{2.537\AA^{-1}}$ respectively
for the Cu-K$\mathrm{\alpha}$ wavelength. In our case, these peaks
cannot be observed due to the presence of the Nb buffer diffraction
peaks, whose intensity is a consequence of the crystalline quality
of the niobium layer. However, it is possible to see an increase in
the background intensity at the $\alpha$-uranium peak positions,
dependent on the thickness of the uranium layers.

The broad hump on the high-angle side of the substrate peak is close
to the bulk bcc (110) iron position and there are no peaks at other
allowed bcc Fe crystallographic directions, suggesting a preferred
orientation in this growth direction. This confirms predictions
considering only the likely growth in the direction of the most
closely packed plane as discussed earlier. The lack of any intensity
at all at iron layer thicknesses of $\mathrm{<20\AA}$ suggests that
this represents a crystalline limit, below which the growth would be
expected to be amorphous and consequently of reduced magnetisation.
The positions of the iron peaks were used to deduce values for the
average lattice spacing in the growth direction, $\mathrm{d_{Fe}}$,
and the mean crystallite size, D, was determined by measuring the
width of the peaks and using the Scherrer equation. This states that
as the diffracting volumes become smaller the peaks broaden, giving
a finite $\mathrm{\Delta\theta}$ width. The size of the diffracting
particles, D, is given by
$\mathrm{D=\frac{K\lambda}{\Delta\theta\cos\theta_{B}}}$, where
K=0.9394 and $\mathrm{\theta_{B}}$ is the Bragg angle
\cite{Patterson}. Values for $\mathrm{d_{Fe}}$ and D, determined by
this method are given in table \ref{UFedifsummary} for a selection
of U/Fe samples.

\begin{table}[htbp]\caption{\label{UFedifsummary} Lattice spacings ($\mathrm{d_{Fe}}$) and particle sizes (D) of Fe layers
for a selection of U/Fe samples, determined by an investigation of
the diffraction peak positions and widths, using the Scherrer
formula. For bulk bcc Fe $\mathrm{d_{(110)}=2.027\AA}$.} \centering
\begin{tabular}{clcc}
\br Sample Number & Composition &
$\mathrm{d_{Fe}}$ ($\mathrm{\AA\pm0.005}$) & D ($\mathrm{\AA\pm2}$) \\
\mr
SN71 & $\mathrm{[U_{9}/Fe_{34}]_{30}}$ & 2.052 & 31.0 \\
SN74 & $\mathrm{[U_{32}/Fe_{27}]_{30}}$ & 2.073 & 23.5 \\
SN75 & $\mathrm{[U_{35.2}/Fe_{27}]_{30}}$ & 2.073 & 23.1 \\
SN76 & $\mathrm{[U_{27.5}/Fe_{57}]_{30}}$ & 2.045 & 49.4 \\ \br
\end{tabular}
\end{table}

The average lattice spacings are larger than the bulk Fe value of
$\mathrm{d=2.027\AA}$, indicating an overall lattice expansion. As
the thickness of the iron layer is increased the lattice spacing
approaches that of the bulk value for a bcc ([110] oriented)
crystal. The particle size scales with Fe layer thickness, but is
several $\mathrm{\AA}$ thinner than the Fe layers. This is
consistent with the picture of a non-coherent growth between the
$\mathrm{\alpha-U}$ and Fe atoms; crystallites which do not extend
across more than one layer due to the poor registry between Fe and U
crystal planes and regions of alloy at the interfaces.

The lower panel of figure \ref{UFeCoxdif} shows a summary of the
X-ray diffraction patterns taken for several U/Co samples. The
changing period of the niobium satellites can be seen on the
low-angle side of the sapphire substrate peak as the buffer
thickness is changed; SN117 has $\mathrm{\sim50\AA}$ Nb and SN118,
$\mathrm{\sim100\AA}$ Nb. It was not possible to see any effect of
varying $\mathrm{t_{U}}$ on the observed diffracted intensity. The
diffraction patterns for the U/Co series of samples are remarkably
similar in character to those of the U/Fe system, since the position
of the hcp (002) cobalt peak lies at almost exactly the same
wavevector as that for bcc (110) iron. The nature of the broad hump
on the high-angle side of the substrate peak is influenced by the
thickness of the cobalt layers and a similar relationship can be
observed between $\mathrm{t_{Co}}$ and the diffracted intensity of
the cobalt layers as was seen for the U/Fe series of samples.

The observed intensity of the cobalt hcp (002) peak, while no other
peaks are observed at other allowed hcp Co crystallographic
directions, indicates a preferred orientation in this growth
direction, expected since it is the most closely packed plane within
the hcp crystal structure. The average particle size, D, and the
lattice spacing, $\mathrm{d_{Co}}$, were determined, using the same
method outlined for the U/Fe samples, and these are summarised in
table \ref{UCodifsummary}.

\begin{table}[htbp]\caption{\label{UCodifsummary}Lattice spacings and particle sizes of Co layers
for a selection of U/Co samples, determined by an investigation of
the diffraction peak positions and widths, using the Scherrer
formula. For bulk hcp Co $\mathrm{d_{(002)}=2.035\AA}$.} \centering
\begin{tabular}{clcc}
\br Sample Number & Composition &
$\mathrm{d_{Co}}$ ($\mathrm{\AA\pm0.005}$) & D ($\mathrm{\AA\pm2}$) \\
\mr
SN116 & $\mathrm{[U_{19}/Co_{42.5}]_{20}}$ & 2.064 & 28.3 \\
SN117 & $\mathrm{[U_{9}/Co_{51}]_{15}}$ & 2.058 & 43.2 \\
SN118 & $\mathrm{[U_{10}/Co_{34.5}]_{20}}$ & 2.075 & 31.7 \\ \br
\end{tabular}
\end{table}

Samples SN117 and SN118 described in figure \ref{UFeCoxdif} (lower
panel) and table \ref{UCodifsummary} were grown at an elevated
substrate temperature of $\mathrm{\sim450K}$. These samples allow
both a layer thickness dependent and a temperature dependent
comparison to be made. For all U/Co samples, the particle sizes
follow a similar trend to that observed for the U/Fe system,
although the crystallite sizes are larger in proportion to the Co
layer thicknesses for the samples grown at elevated temperature. As
for the U/Fe samples, the lattice spacings are expanded compared to
the bulk, but tend towards the bulk value as $\mathrm{t_{Co}}$
increases.

Figure \ref{UGd1xdif} shows a summary of the X-ray diffraction
patterns taken for a series of U/Gd samples. There are a number of
striking differences in the form of the diffracted intensity between
the U/TM metal and the U/Gd multilayers. The multilayer diffraction
peaks occur on the low-Q side of the sapphire substrate peak and
their intensity reaches values up to one tenth of the intensity of
the substrate peak, more than two orders of magnitude larger than
the intensity observed for the U/Fe and U/Co systems. This indicates
a far greater degree of crystallinity for U/Gd than for U/TM
samples. The diffraction satellites from the highly crystalline
niobium buffer layers are not observable in most cases above the
multilayer diffraction peaks, although a contribution from the
niobium can be observed as a shoulder on the low angle side of the
substrate peak. A gadolinium film (SN62) of $\mathrm{\sim500\AA}$
was grown to confirm the expected position of the diffraction peaks
in the multilayer samples and diffraction data for this sample are
shown as the dashed curve in figure \ref{UGd1xdif}.

The series of U/Gd multilayers of figure \ref{UGd1xdif} was grown to
investigate the relationship between $\mathrm{t_{Gd}}$ and
$\mathrm{t_{U}}$ on the structural and magnetic properties of the
U/Gd system. The accepted values for the lattice parameters of the
hexagonal close-packed crystal structure of gadolinium are
$\mathrm{a=3.631\AA}$ and $\mathrm{c=5.777\AA}$, giving a
contraction from the hard sphere model for the c/a ratio (1.633) to
1.591. In these measurements we are sensitive only to length scales
in the z-axis direction, perpendicular to the plane of the film, so
our discussion will centre around the c-axis lattice parameter and
the lattice spacings.

In the case of the single film of gadolinium the (002) peak is
centred at a Q value of $\mathrm{2.152\AA^{-1}}$, corresponding to a
c-axis lattice parameter of $\mathrm{5.840\AA}$, an expansion from
the bulk of about 1\%. It is also possible to observe intensity from
the niobium buffer at $\mathrm{2.450\AA^{-1}}$ and a peak at
$\mathrm{2.040\AA^{-1}}$, corresponding to the hcp (100) reflection
that occurs in the bulk at $\mathrm{1.998\AA^{-1}}$. This shift to
higher Q in the thin Gd film suggests a contraction of the lattice
along the basal plane of $\mathrm{\sim2\%}$ towards a value of
$\mathrm{a=3.556\AA}$ and a c/a ratio of $\mathrm{1.642}$.

Indicated by a dashed arrow on figure \ref{UGd1xdif}, as
$\mathrm{t_{Gd}}$ increases there is a distinct increase in
intensity of one of the component peaks in the diffraction patterns,
close to the hcp (002) peak observed for the thin Gd film. This
increase in intensity is accompanied by a shift in position from the
low angle side of the (002) peak towards the thin film value,
indicating a lattice expansion for thinner Gd layers.

As the uranium layer thickness, $\mathrm{t_{U}}$, is varied there is
a clearly visible increase in the intensity of one of the component
peaks in the X-ray diffraction spectra, at $\mathrm{2.245\AA^{-1}}$.
This peak does not relate to any of the known peak positions in the
$\alpha$-U phase, but could correspond to the (002) peak of an [001]
preferred orientation hcp-U crystal structure. Recent theoretical
and experimental evidence \cite{BerbilBautista, Molodtsov} supports
the existence of a stable hcp-U phase established in thin film
structures, for an uranium film grown on a [110] oriented bcc,
tungsten single crystal substrate. STM images \cite{BerbilBautista}
have described a hexagonal arrangement of atoms with a U-U distance
of $\mathrm{a=3.5\pm0.5\AA}$, although a previous report by
Molodtsov et al. \cite{Molodtsov} suggested a U-U distance of
$\mathrm{3.2\pm0.5\AA}$. A theoretical model \cite{BerbilBautista},
employing the local density approximation (LDA) supports the idea
that an hcp-U crystal structure can be stabilised with a c/a axis
ratio of 1.8, appreciably larger than the hard sphere, hcp model
value of 1.633. The predicted values for c and a are
$\mathrm{5.35\AA}$ and $\mathrm{2.97\AA}$, respectively. However, it
is accepted that there is a tendency for the LDA theory to
over-compress the lattice and the actual values for the c and a axis
parameters may be larger than these values.

Assuming the uranium stacks along the [001] axis, a reflection would
be observed in the diffraction spectrum at Q
$\mathrm{\sim2.3\AA^{-1}}$, which is close to the position of the
diffraction peak attributed to the uranium in figure \ref{UGd1xdif}.
Moreover, this peak position results in a lattice spacing along the
c-axis only $\mathrm{5\%}$ larger than that expected for gadolinium,
which could provide the mechanism for the growth and orientation of
the exotic hcp (001) phase of uranium.

Figure \ref{UGd2xdif} shows a summary of the X-ray diffraction
patterns taken for a series of U/Gd multilayers grown to investigate
samples with thin layers, $\mathrm{<20\AA}$, and to observe the
effect of increased substrate temperature on the crystalline
structure. The thinnest of these films (SN122) had a bilayer
thickness of just $\mathrm{22.5\AA}$ yet an appreciable diffracted
intensity was still observable. It is clear that the crystalline
limit for this multilayer system exists only for very thin layers.
All of the samples shown in figure \ref{UGd2xdif} consist of thin
uranium layers of approximately $\mathrm{11\AA}$. The prominent peak
in these diffraction spectra occurs at the hcp-U peak position
observed previously, which shifts to lower Q as the gadolinium layer
thickness increases. This indicates that the uranium grows in a more
crystalline manner than the gadolinium layers at low values of
$\mathrm{t_{U}}$ and $\mathrm{t_{Gd}}$ respectively, and that as
$\mathrm{t_{Gd}}$ increases the average U lattice spacing is
increased, possibly as a consequence of the Gd crystallinity.
Samples SN123 and SN124 share similar compositions, but were grown
at room temperature and an elevated substrate temperature of 600K
respectively. The diffracted intensity of the latter shares the same
characteristics as that grown at room temperature, but is more than
an order of magnitude greater, indicating a more crystalline
assembly at elevated temperatures. Recalling the X-ray reflectivity
from these two samples (see figure \ref{UGdxref} (d)) along with the
information taken from the high-angle diffraction measurements it is
reasonable to infer that the relative amount of interdiffusion
between the U and Gd species is small even at elevated growth
temperatures. Remembering also the summary of the microstructural
growth properties for sputtered films \cite{Thornton}, as the
substrate temperature is increased the ratio $\mathrm{T/T_{m}}$
increases, where $\mathrm{T_{m}}$ is the melting point of the
respective elements, suggestive of a more columnar crystal growth,
which could be responsible for the large rms roughness needed to
provide the rapid decay of intensity observed in the X-ray
reflectivity spectrum for sample SN124.

\subsubsection{Discussion}

\begin{figure}[htbp]
\center
\mbox{\subfigure{\epsfig{figure=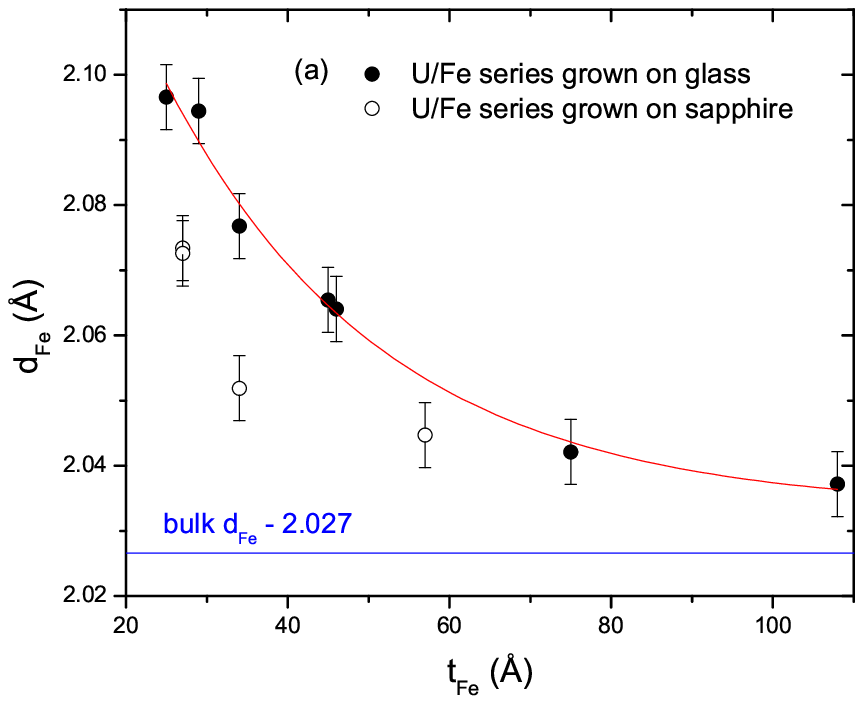,width=0.50\textwidth,bb=10
10 270 225,clip}}\quad
\subfigure{\epsfig{figure=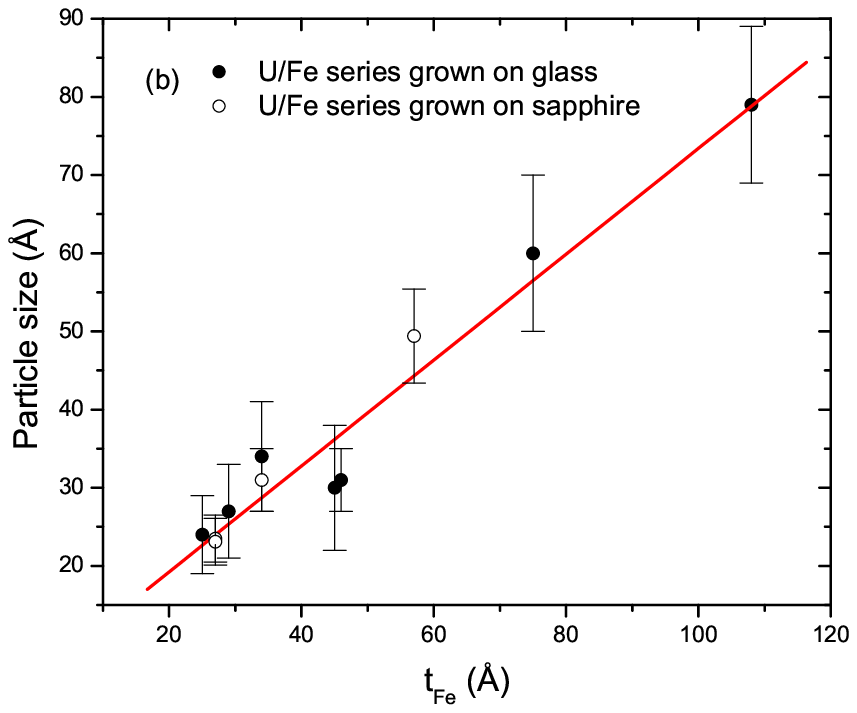,width=0.50\textwidth,bb=10 10
270 225,clip}}}\caption{\label{UFecomparison}Comparison of (a) the
iron lattice spacings, $\mathrm{d_{Fe}}$, and (b) particle sizes for
U/Fe samples grown on glass \cite{Beesley1, Beesley2} and those
grown on sapphire (present work).}
\end{figure}

The U/Fe series of samples considered within this study can be
compared directly to results published previously
\cite{Beesley1,Beesley2}, to investigate the differences and
similarities in the growth of the two sets of multilayers. Figure
\ref{UFecomparison} (a) compares the lattice spacings,
$\mathrm{d_{Fe}}$, and figure \ref{UFecomparison} (b), the particle
sizes taken from the structural characterisation results published
by Beesley et al. \cite{Beesley1} with those obtained more recently
for the samples grown on sapphire substrates with niobium buffer and
capping layers. The particle sizes of the two sets of samples follow
the same trend, tending to increase with increasing Fe layer
thickness. The lattice parameter, however, seems to follow a much
steeper exponential trend towards the bulk value for large
$\mathrm{t_{Fe}}$ in the case of samples grown on sapphire than for
those grown on glass.

The U/Gd diffraction spectra are markedly different from those
observed for the U/TM multilayers investigated thus far. The peaks
attributed to the hcp Gd (002) and hcp U (002) crystalline phases
exhibit a mismatch of $\mathrm{<5\%}$, resulting in the presence of
satellite peaks either side of these primary ones. This observation
is a direct consequence of a coherent growth between respective U
and Gd layers, giving coherent scattering from crystalline planes in
many layers.

\begin{table}[htbp]\caption{\label{bilayer}Comparison of the
bilayer thicknesses for (a) samples with increasing
$\mathrm{t_{Gd}}$ and (b) samples with increasing $\mathrm{t_{U}}$,
determined by both high ($\mathrm{t_{Bdif}}$) and low angle
($\mathrm{t_{Bref}}$) X-ray diffraction. $\mathrm{\bar{x}}$
represents the average separation of the high-angle diffraction
satellites in $\mathrm{\AA^{-1}}$.} \centering
\subtable[]{\begin{tabular}{ccrr}\br $\mathrm{t_{Gd}}$ ($\mathrm{\AA\pm2}$) & $\mathrm{\bar{x}}$ ($\mathrm{\AA^{-1}}$) & $\mathrm{t_{Bdif}}$ ($\mathrm{\AA\pm2}$) & $\mathrm{t_{Bref}}$ ($\mathrm{\AA\pm2}$) \\
\mr
33.0 & 0.122 & 51.5 & 59 \\
54.0 & 0.087 & 72.2 & 80 \\
76.0 & 0.067 & 93.8 & 102 \\
\br\end{tabular}}\space\space
\subtable[]{\begin{tabular}{ccrr}\br $\mathrm{t_{U}}$ ($\mathrm{\AA\pm2}$) & $\mathrm{\bar{x}}$ ($\mathrm{\AA^{-1}}$) & $\mathrm{t_{Bdif}}$ ($\mathrm{\AA\pm2}$) & $\mathrm{t_{Bref}}$ ($\mathrm{\AA\pm2}$) \\
\mr
39.0 & 0.110 & 57.1 & 59.0 \\
63.5 & 0.078 & 80.6 & 83.5 \\
89.0 & 0.059 & 106.5 & 109.0 \\
\br\end{tabular}}
\end{table}

Table \ref{bilayer} compares the bilayer thicknesses determined by
both high- and low-angle X-ray diffraction methods for a series of
U/Gd samples. For the samples with thick Gd layers, the values of
the bilayer thickness determined by the separation of the high-angle
diffraction satellites are consistently $\mathrm{\sim8\AA}$ lower
than those indicated by X-ray reflectivity measurements, whereas for
samples with thick U layers the values are only
$\mathrm{\sim2.5\AA}$ lower. This result suggests that a small
region of the bilayer is not crystalline (likely to be present at
the interface), and that this noncrystalline component is larger in
samples with thick Gd layers than in those with large
$\mathrm{t_{U}}$. This result is consistent with the variation in
roughness in the X-ray reflectivity measurements.

\begin{figure}[htbp]
\centering
\includegraphics[width=0.7\textwidth,bb=10 10 225 245,clip]{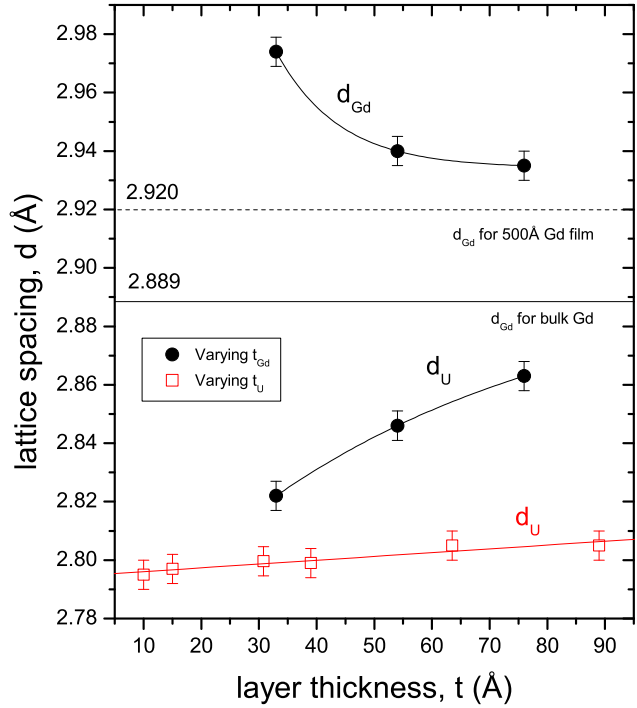}\caption
{\label{UGddiffraction_summary}Variations in the lattice spacings of
uranium and gadolinium as a function of $\mathrm{t_Gd}$ (full
points) and $\mathrm{t_{U}}$ (open squares). Values for the bulk and
thick Gd film gadolinium lattice parameter are labeled. Solid lines are guides for the eye.}
\end{figure}

When the layers grow coherently in a multilayer it is possible for
variations in the layer thickness of one element to affect the
strain profile of another, which will be reflected in the lattice
spacing values. Figure \ref{UGddiffraction_summary} summarises
the lattice spacing values as determined from the X-ray diffraction
profiles for a number of U/Gd samples. The variation of both
$\mathrm{d_{Gd}}$ and $\mathrm{d_{U}}$ are shown as a function of
$\mathrm{t_{Gd}}$, full points (black), the dependence of the
uranium lattice parameter upon the U layer thickness is represented
by the open squares (red). It was not possible to distinguish the
gadolinium diffraction peak positions in the case of varying
$\mathrm{t_{U}}$, since the gadolinium layers were too thin to give
an appreciable diffracted intensity.

As described earlier, the lattice parameter in a sputtered thin film
of gadolinium ($\mathrm{2.920\AA}$) is expanded compared to that of
bulk Gd ($\mathrm{2.899\AA}$); both of these values are clearly
marked in figure \ref{UGddiffraction_summary}. For multilayers
containing thin gadolinium layers the Gd lattice is further
expanded, but contracts towards the sputtered thin film value as the
layers become thicker. There is very little observable change in the
U lattice spacing, $\mathrm{d_{U}}$, as $\mathrm{t_{U}}$ is varied.
For the case of these samples the gadolinium layer thickness is
constant at $\mathrm{20\AA}$. However, a slight expansion of the
lattice is observable for thick U layers. An interesting result
observed here is the dependence of $\mathrm{d_{U}}$ upon the
gadolinium layer thickness, with $\mathrm{t_{U}\sim26\AA}$. A
consideration of the lattice parameter sizes of the hcp (001) Gd and
hcp (001) U phases, reveals a likely strain acting to expand the U
lattice. The trend observed in figure \ref{UGddiffraction_summary}
implies an increase in the strain acting on the U layers, as
$\mathrm{t_{Gd}}$ is increased, which provides a mechanism for the
observed increase in $\mathrm{d_{U}}$.

\section{Conclusions}

The X-ray reflectivity spectra of all samples display well-defined
Bragg peaks, which give accurate determinations of the bilayer
thickness. The thicknesses of the individual layers are then
obtained by maintaining consistency across a range of measurements.
The inclusion of niobium buffer and capping layers has considerably
reduced the complexity of the structural model required to simulate
the reflectivity spectra of previous samples \cite{Beesley1}.

No significant improvement was observed in the crystalline growth of
U/Fe samples on sapphire substrates with a Nb buffer layer when
compared with those grown on glass \cite{Beesley1}. The poor
crystalline quality arises from the large mismatches in atomic sizes
and lattice spacings at the U/Fe interfaces and chemical diffusion
processes; this situation is replicated in the U/Co system. The
similarities between the U/transition metal systems stem from the
similar atomic volumes, resulting in no significant diffracted
intensity observable from the U layers in these systems. The
interdiffusion at the interfaces, causing an alloy region,
observable as a nonmagnetic Fe component in earlier M\"{o}ssbauer
spectroscopy data \cite{Beesley2} is likely to be occur also in the
U/Co system. The further determination of an amorphous (reduced
moment) ferromagnetic component from this data explains the lack of
diffracted intensity for U/Fe multilayers with $\mathrm{t_{Fe} <
20\AA}$. The proposed structure for the ferromagnetic layers in
U/transition metal systems is then:
$\mathrm{UT_{alloy}|T_{amorphous}|T_{crystalline}|TU_{alloy}}$,
where T represents the transition metals iron and cobalt.

The situation in the case of the U/Gd system is considerably
different. Although an atomic volume mismatch in one dimension of
$\mathrm{\sim14\%}$ is still present between U and Gd, the strain is
in the opposite sense to that present in U/Fe andU/Co multilayers.
The outcome is the observation of an intense diffraction peak that
does not correspond to any known for $\alpha-U$, but at a position
close to that reported for a novel hcp phase of uranium
\cite{BerbilBautista}. The lattice spacing, $\mathrm{d_{U}}$,
determined by this peak for thick U layers is about
$\mathrm{2.80\AA}$ and does not change significantly as the U layer
thickness is varied (for constant $\mathrm{t_{Gd}}$ of
$\mathrm{20\AA}$). This gives a c-axis lattice parameter for hcp
uranium of 5.60\AA, somewhat larger than values put forward in the
study of uranium grown on tungsten \cite{BerbilBautista}. Assuming
the same atomic volume of $\mathrm{20.5\AA^{3}}$ for hcp U as that
for $\mathrm{\alpha-U}$ and taking the c-axis parameter determined
from X-ray diffraction measurements, the resulting a-axis value is
$\mathrm{2.91\AA}$, giving a c/a ratio close to 1.9, much larger
than that expected from a hard sphere model of the crystal
structure. The lattice parameter of the a-axis also represents the
U-U nearest neighbour distance in the hcp crystal structure, which
in the $\mathrm{\alpha-U}$ phase is $\mathrm{2.84\AA}$. A comparison
of the local environments of U atoms in these two structures reveals
a considerable change in the coordination and a lattice expansion of
$\mathrm{\sim2.5\%}$. The observation of such intense diffraction
spectra is then remarkable, considering the likely in-plane lattice
spacings. We hope to verify these findings directly with further
experiments. The mismatch, between the Gd ($\mathrm{a=3.56\AA}$ for
the sputtered film) and hcp U ($\mathrm{a\sim2.91\AA}$) is
$\mathrm{\sim22\%}$, yet growth along the common c-axis remains
good. This result implies that considerable strains are present at
the U/Gd interfaces, extending perhaps to a considerable distance
into both layers.

The mismatch between the Gd hcp (001) and U hcp (001) lattice
parameters along the c-axis is less than 5\%, which explains the
observation of satellite diffraction peaks, produced by coherent
scattering from crystalline planes in many different layers. The
bilayer thickness values determined from the separation of the
satellite peaks were several {\aa}ngstr\"{o}m less than those
determined by X-ray reflectivity, indicating a small interface
region of noncrystalline material. A comparison of the U/Gd and the
U/Co systems, suggests that such crystalline growth of the U layers
in the U/Gd system is not simply due to the hexagonal packing
arrangement of the gadolinium atoms, since the Co layers also adopt
the hcp (002) crystal structure, but is due to the larger Gd atomic
volume and possibly also to the different electronic configurations
of the element.

X-ray reflectivity measurements of U/Gd multilayers revealed a
strong dependence of the average roughness per bilayer upon the
gadolinium layer thickness, suggesting a step-like roughness of the
crystalline gadolinium layers, due to a columnar growth mechanism.
This idea is further supported by the observation of increased
intensity in the case of the X-ray diffraction spectra in figure
\ref{UGd2xdif} for a sample grown at an elevated substrate
temperature, indicating a better degree of crystallinity within the
layers rather than an increase in the rate of interdiffusion. The
X-ray reflectivity spectra shown in figure \ref{UGdxref} (d) then
reveal a much larger degree of roughness present in the sample grown
at elevated temperature, which can be related to a columnar-like
growth mechanism \cite{Thornton}.

The magnetic properties of U/Fe, U/Co and U/Gd multilayers are
addressed in paper II of this series. Future measurements are
planned to investigate the unusual growth mechanisms and structures
of the U/Gd system, particularly the novel hcp U structure.

\ack

RS acknowledges the receipt of an EPSRC research studentship.

\section{References}
\bibliographystyle{unsrt}
\bibliography{Umultilayers1}
\end{document}